# Differential measurement of atmospheric refraction with a telescope with double fields of view [*]

Yong Yu[1], Jian-Jun Cao[1], Zheng-Hong Tang[1], Hao Luo[1,2], Ming Zhao[1]

[1] Shanghai Astronomical Observatory, Chinese Academy of Sciences, Shanghai 200030, China;
   *yuy@shao.ac.cn*

[2] University of Chinese Academy of Sciences, Beijing 100049, China

Abstract For the sake of complete theoretical research of atmospheric refraction, the atmospheric refraction under the condition of lower angles of elevation is still worthy to be analyzed and explored. In some engineering applications, the objects with larger zenith distance must be observed sometimes. Carrying out observational research of the atmospheric refraction at lower angles of elevation has an important significance. It has been considered difficult to measure the atmospheric refraction at lower angles of elevation. A new idea for determining atmospheric refraction by utilizing differential measurement with double fields of view is proposed. Taking the observational principle of HIPPARCOS satellite as a reference, a schematic prototype with double fields of view was developed. In August of 2013, experimental observations were carried out and the atmospheric refractions at lower angles of elevation can be obtained by the schematic prototype. The measured value of the atmospheric refraction at the zenith distance of 78.8 degree is 240.23" ± 0.27", and the feasibility of differential measurement of atmospheric refraction with double fields of view was justified. The limitations of the schematic prototype such as inadequate ability of gathering light, lack of accurate meteorological data recording and lower automatic level of observation and data processing were also pointed out, which need to be improved in subsequent work.

## 1 INTRODUCTION

Atmospheric refraction is the difference between the direction before the light of a celestial body enters into the atmosphere and the direction when it reaches the observer. In comparison with the other factors which affect the direction of the celestial body, atmospheric refraction is characterized by its uncertainty, mainly in: the atmospheric refraction at different locations could be different; at the same location, the atmospheric refractions of different directions are not exactly the same. Especially at larger zenith distance, the performance of the two cases above is more prominent. One of the important issues of ground-based astrometry is how to establish more reasonable atmospheric refraction models and how to improve the correction accuracy of atmospheric refraction.

[*] Supported by the National Natural Science Foundation of China.

With the continuous release of high-precision star catalogs, the method of relative celestial positioning based on photographic technique can effectively reduce the influence of atmospheric refraction. But in many areas, the high-precision correction of atmospheric refraction is still necessary. For example, in the process of creating a global pointing model for a telescope, the stars at lower angles of elevation must be observed so that the atmospheric refraction at larger zenith distance requires to be known precisely. As another example, when tracking a low Earth orbit target using the method of shafting positioning, or when a space tracking telemetering and command ship works at sea, due to the restrict of tracking conditions or special requirement of monitoring and control task, the situation of large zenith distance will be often encountered. When carrying out the trajectory measurement in shooting ranges, the farther the target is away from the observer, the lower angle of elevation it becomes, and even a negative angle of elevation will appear, the correction error of atmospheric refraction has become one of the main errors for the measurement in shooting ranges (Wang et al. (2013); Zhou & Zhao (2012)). The usual method of correcting atmospheric refraction is to adopt a general theoretical model (Zhao (2012); Zhang et al. (2013)), but atmospheric refraction has local characteristics, especially at large zenith distance, the actual value of the atmospheric refraction may deviate seriously from the theoretical one. Therefore, how to obtain high-precision atmospheric refraction at large zenith distance is a key of improving the accuracy of global telescope pointing model, heightening the observational accuracy and prolonging the observational arcs of the tracked target. In addition, in the field of space geodesy, atmospheric refraction delay is getting more and more attention. There necessarily exists a sort of internal relation between atmospheric refraction and atmospheric refraction delay with the atmospheric refractive index as the link. Utilizing this kind of internal relation, some researchers proposed the method for transforming atmospheric refraction observational model into atmospheric refraction delay correction model, so as to overcome the shortages of the adopted theoretical model and empirical model (Mao et al. (2006); Mao et al. (2009)). This sort of application not only inputs the renewed vitality into the observational research on atmospheric refraction, but sets an even higher demand on it.

The fundamental way of improving the correction accuracy of atmospheric refraction is adopting an effective method to carry out actual measurement at the observing station, combining the instantaneous meteorological data and building the position-dependent observational model of atmospheric refraction. It has been considered difficult to measure the atmospheric refraction at lower angles of elevation. In 2008, we proposed a set of differential measurement method of atmospheric refraction (Yu et al. (2009)). The principle can be outlined as: a telescope with a larger field of view is employed to make a series of observations of the starry sky at different angles of elevation, the derivatives of various orders of atmospheric refraction function at different zenith distances are calculated according to the comparison between observational and theoretical arcs of the constellation in each field of view, and finally the actually observed values of atmospheric refraction can be found via numerical integration. Different from the absolute measurement methods before, such as the determination of local atmospheric refraction carried out with a reflecting low latitude meridian circle (Mao et al. (2009)), this method could weaken the effect of the systematic errors like the local parameters and instrumental parameters. Several test observations have been done and indicate that the idea of the determination of atmospheric refraction with the differential method is feasible, but the observational results also show that the measurement using the telescope with single field of view would be affected by cumulative error, which will influent the final observed values of atmospheric refraction.

How to improve the observational accuracy is the key problem for the differential measurement of atmospheric refraction. After analysis and study, we proposed an idea of differential measurement of atmospheric refraction with a telescope with double fields of view, developed a schematic prototype, and conducted experimental observations based on the prototype to investigate its feasibility. In the present article, the principle of differential measurement of atmospheric refraction with double fields of view is introduced in the 2nd section, the information about the schematic prototype is shown in the 3rd section, the experimental observation results are given in the 4th section, and finally the concluding remarks are given.

## 2 FUNDAMENTAL PRINCIPLE OF DIFFERENTIAL MEASUREMENT OF ATMOSPHERIC REFRACTION WITH DOUBLE FIELDS OF VIEW

A telescope with double fields of view is employed to observe two sky regions simultaneously at different zenith distances, the atmospheric refraction at larger zenith distance is calculated according to the comparison between actual arc length of the centers of the two fields of view (determined by measurement instrument itself) and theoretical one (calculated by apparent positions of observed stars). The specific calculation can be described as:

Let $z_i$ be the true zenith distance of a star and $z_i'$ is the observed zenith distance, then

$$\Delta z_i = z_i - z_i' \tag{1}$$

is the atmospheric refraction. For the simultaneous observations of the two fields of view at different zenith distances of the same azimuth, there is

$$L_0 = (z_2 - \Delta z_2) - (z_1 - \Delta z_1) = (z_2 - z_1) - (\Delta z_2 - \Delta z_1) \tag{2}$$

where $z_1$, $\Delta z_1$ and $z_2$, $\Delta z_2$ represent the corresponding variables of small and large zenith distances; $L_0$ is determined by the measurement instrument itself, which represents the actual angle distance between the two fields of view; $z_2 - z_1$ is the difference of the true zenith distances between the two fields of view, which can be calculated from catalog positions of stars, observation time and the local constants of the observing site; $\Delta z_1$ is the atmospheric refraction at small zenith distance, and it can be obtained by the theoretical model which is accurate enough under the condition of smaller zenith distance. So the atmospheric refraction $\Delta z_2$ at large zenith distance can be obtained according to the equation (2), that is

$$\Delta z_2 = (z_2 - z_1) - L_0 + \Delta z_1 \tag{3}$$

But in fact, it is impossible to observe the two points exactly at the same azimuth because of the pointing error of the telescope, i.e. the centers of the two fields of view cannot be on the same vertical circle. There will be some difference in azimuth between the two sky regions, so the calculation of spherical triangle need to be carried out to get the atmospheric refraction at large zenith distance. As shown in figure 1, let $Z$ be the zenith, $\sigma_0$ and $\sigma_1$ be the observed positions of the centers of the two fields of view, their corresponding horizon coordinates are $(A_0, h_0)$ and $(A_1, h_1)$, and there is

$$\sigma_0\sigma_1 = L_0 \tag{4}$$

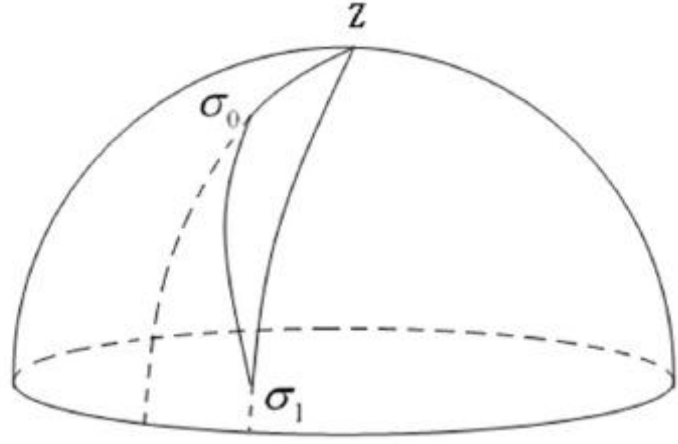

Fig. 1  Celestial spherical schematic of the observation with double fields of view.

Their azimuths and true zenith distances can be calculated according to the catalog positions of stars, observation time and local constants of the observing site. Under normal circumstances, atmospheric refraction does not affect the azimuth of a celestial body, so there is

$$\angle\,\sigma_0 Z \sigma_1 = \Delta A = A_0 - A_1 \tag{5}$$

$Z\sigma_0$ can be calculated by the theoretical model of atmospheric refraction. In the spherical triangle of $\angle\,Z\sigma_0\sigma_1$ , sine formula of spherical triangle expressed as

$$\frac{sinZ\sigma_0}{sin\angle\,Z\sigma_1\sigma_0} = \frac{sin\sigma_0\sigma_1}{sin\Delta A}$$

can be used to get $\angle\,Z\sigma_1\sigma_0$ at first, and then cosine formula of spherical triangle expressed as

$$cos\sigma_0\sigma_1 = cosZ\sigma_0 cosZ\sigma_1 + sinZ\sigma_0 sinZ\sigma_1 cos\Delta A \tag{7}$$

is combined with five-element formula of spherical triangle expressed as

$$sinZ\sigma_0 cos\Delta A = cos\sigma_0\sigma_1 sinZ\sigma_1 - sin\sigma_0\sigma_1 cosZ\sigma_1 cos\angle\,Z\sigma_1\sigma_0 \tag{8}$$

to get $Z\sigma_1$ , which is the observed zenith distance of $\sigma_1$. Finally, the atmospheric refraction at the large zenith distance can be obtained according to the difference between the observed and the true zenith distance.

## 3  SCHEMATIC PROTOTYPE OF TELESCOPE WITH DOUBLE FIELDS OF VIEW

In order to justify the feasibility of differential measurement of atmospheric refraction with double fields of view, a telescope that can observe two sky regions simultaneously is needed. The telescope must meet the critical requirement that the angle distance of the two observing sky regions should be fixed. A simple and direct simple idea is that the two sky regions can be observed simultaneous by a binocular telescope. But under the influence of the gravity and ambient temperature change, the tubes and the bracket of the binocular telescope may produce different deformation inevitably in the process of measurement, which will change the angle distance between the two fields of view. With regard to this, the observational principle of HIPPARCOS satellite was referred (Ratier & Batut (1989)). The telescope of HIPPARCOS satellite is an all-reflecting Schmidt one and consists of three mirrors: a beam combiner, a flat-folding mirror and a spherical primary mirror, as shown in figure 2. The beam combiner is made from a polished zerodur blank, cut into two halves and reassembled with a wedge angle of 29 degree. The star light from two sky regions with the angle of 58 degree can be reflected to the flat-folding mirror through the beam combiner, and then converged to the focal plane through the spherical primary mirror, which makes it possible to measure the angle distances between the stars from the two sky regions.

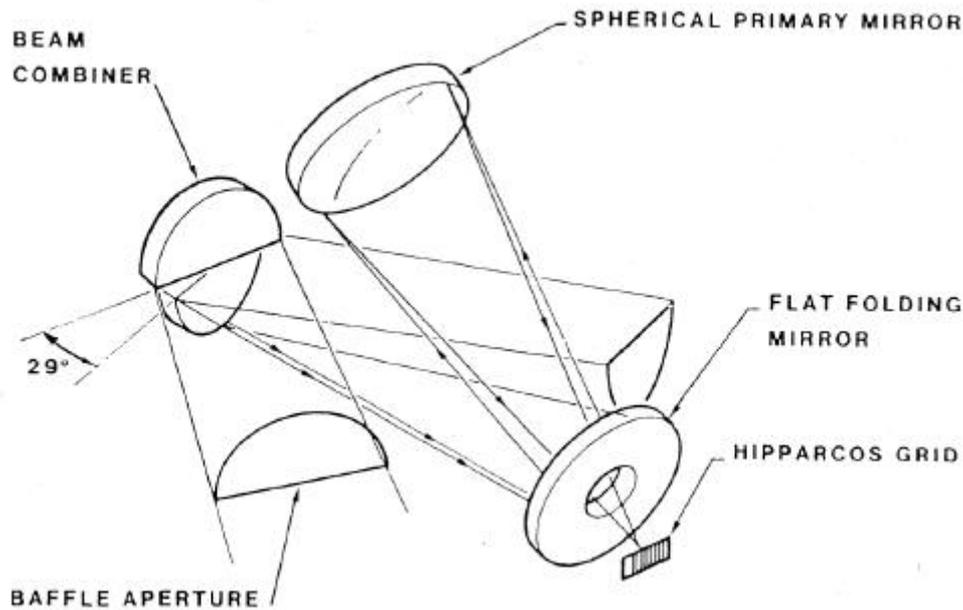

Fig. 2  Configuration of the telescope of the HIPPARCOS satellite (Ratier & Batut (1989)).

Based on the observational principle of HIPPARCOS satellite, a double-sided reflective device (we call this angle reflector) similar to the beam combiner is supposed to install in front of a telescope tube, which can reflect the star light from two different directions into one tube, as shown in figure 3. At this point, the angle distance of the two observing sky regions is determined only by the angle between two reflection surfaces, which can avoid the change caused by the deformation of tubes or bracket. A whole piece of zerodur blank is cut, polished and coated in order to ensure the solidity of the reflection angle. The angle of the two reflection surfaces is designed as 30 degree, as shown in figure 4, which can reflect the star light of the two sky regions with the angle of 60 degree to one direction.

When carrying out the measurement of atmospheric refraction, first the telescope is pointed to the zenith to make the star light from two sky regions near the zenith (their zenith distances are both 30 degree or so) image on CCD camera. Since the theoretical model value of atmospheric refraction near the zenith is accurate enough (better than 0.1 arc second), the actual reflection angle of the reflector can be determined by the observed positions of the two fields of view centers. Then, the telescope is pointed to an elevation, e.g. the angle of elevation of 30 degree, thus the stars of the zenith distance of 30 degree and the stars near horizon can be imaged on CCD camera. The atmospheric refraction near horizon can be calculated according to the principle described in the $2^{nd}$ section.

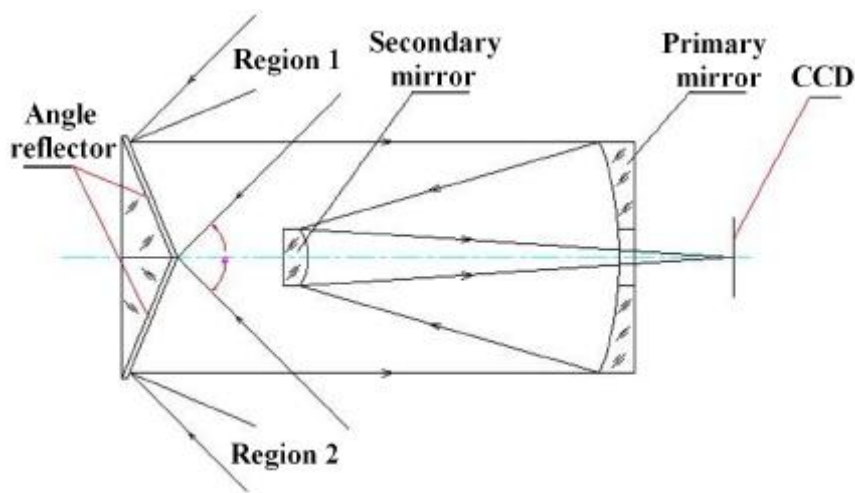

Fig. 3  Configuration of the telescope with double fields of view for atmospheric refraction.

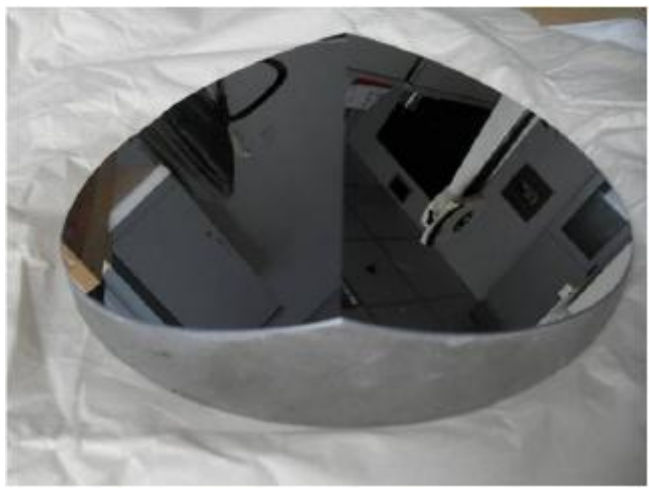

Fig. 4  Photo of the angle reflector.

2012-2013, we have developed a schematic prototype of atmospheric refraction with double fields of view by the modification on a Maksutov tube. The main parameters of the schematic prototype are listed in Table 1 and the photo is shown in figure 5.

## 4  RESULTS OF EXPERIMENTAL OBSERVATION

On August 9, 2013, we carried out an experimental observation with the schematic prototype at Anji station of Shanghai Astronomical Observatory. The station locates at the longitude of 119.5976333°, the latitude of 0.4694055° and the altitude of 943.0m. The temperature was 24.3℃ and the barometric pressure was

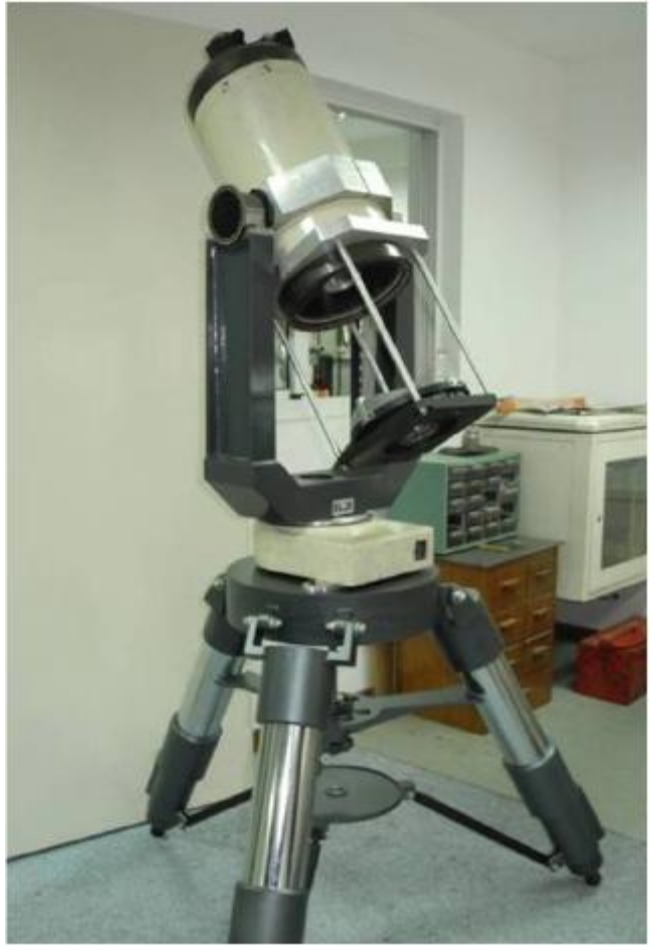

Fig. 5  Photo of the schematic prototype of atmospheric refraction with double fields of view.

Table 1  Main parameters of the schematic prototype of atmospheric refraction with double fields of view

| Telescope tube type | Maksutov |
|---|---|
| Tube diameter | 20cm |
| Tube focal length | 2m |
| Bottom diameter of reflector | 20cm |
| Angle of reflection surfaces of reflector | $30''$ |
| CCD model | Apogee U9000 |
| CCD pixel size | $12\mu m$ |
| CCD array | 3056pixel $\times$ 3056pixel |
| Field of view | $1.0'' \times 1.0''$ |
| Pixel scale of CCD | $2.5''$/pixel@bin2 |
| Operation method | Manual |

906 mba. First, the telescope was pointed to the zenith to determine the reflection angle of the reflector. Then, the telescope was pointed to six directions of different angles of elevation near the azimuth of North and get the measured values of the atmospheric refraction at the observed zenith distance of $49.4''$,$63.9''$, $69.0''$, $74.1''$ and $78.8''$. To measure the atmospheric refraction at low elevation, due to weak support of the heavy reflector in figure 5, the incident light to telescope is inevitably flexed or deflected slightly from the expected direction. However, it doesn't matter to the refraction calculation, which is determined by star positioning in the double fields with a fixed angle.

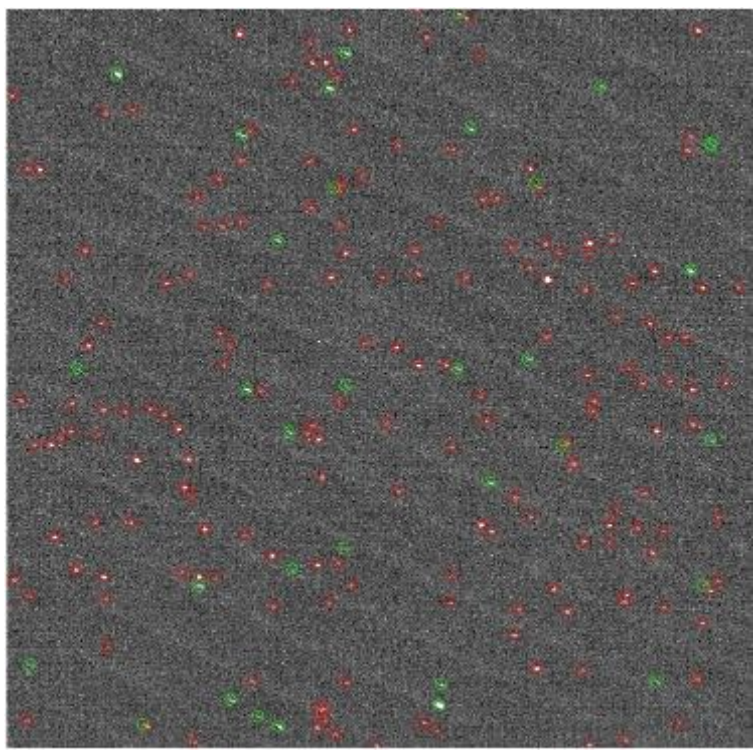

Fig. 6  Example of the CCD observation image in the case of pointing the telescope to the zenith.

## 4.1  Determination of the reflection angle of the reflector

A total of 15 observations were performed to determine the reflection angle of the reflector, and Figure 6 shows an example of the CCD observation image, where the stars marked with green circles and red circles were from north and south sky region of the zenith distances of 30 degree.

When carrying out observations with double fields of view, the stars from two sky regions were imaged simultaneously. As the moving velocities and directions of the stars from two sky regions are different, the function of telescope tracking stars cannot be used, which makes the efficient exposure time of star determined by the residence time of star image on one CCD pixel. As the moving velocity of the stars from south sky region is faster, their efficient exposure time is shorter so that the number of observed stars from south sky region is less than that from north sky region. Table 3 lists the measured value of the reflection angle according to each observation image and the average is $59.4646413'' \pm 0.09''$.

## 4.2  Measured result of atmospheric refraction

The diameter of the Maksutov tube used in schematic prototype is smaller and its focal length is longer. Under the influence of atmospheric distinction, it is impossible for the schematic prototype to observe sufficient stars at any time at the sky region with lower angle of elevation. In order to justify the feasibility of the principle, only north sky regions were observed experimentally, where stars move slowly and stars' efficient exposure time is relatively long. The telescope was pointed successively to the zenith distance of 20, 35, 40, 45 and 50 degree respectively to get the measured values of atmospheric refraction at the zenith distance of 49.4, 63.9, 69.0, 74.1 and 78.8 degree. The results are listed in Table 3, which also gives the comparison results with Pulkovo atmospheric refraction table for reference.

Table 2 Measured value of the reflection angle according to each observation image

| No. | Measured value of reflection angle |
|---|---|
| 1 | 59.4647446″ |
| 2 | 59.4646272″ |
| 3 | 59.4646842″ |
| 4 | 59.4645428″ |
| 5 | 59.4646241″ |
| 6 | 59.4646586″ |
| 7 | 59.4648003″ |
| 8 | 59.4645161″ |
| 9 | 59.4646490″ |
| 10 | 59.4646991″ |
| 11 | 59.4644882″ |
| 12 | 59.4646314″ |
| 13 | 59.4646355″ |
| 14 | 59.4645056″ |
| 15 | 59.4648016″ |
| Average | 59.4646413″ ± 0.09″ |

It can be seen from table 3 that the atmospheric refraction at larger zenith distance can be obtained by the schematic prototype and the average value at the observed zenith distance of 78.8 degree is 240.23″±0.27″. The experiment indicates that differential measurement of atmospheric refraction with a telescope with double fields of view is feasible. From the comparison with Pulkovo table, it can be seen that the discrepancy between the measured values and the tabular values increases gradually with the increase in the observed zenith distance. The reasons possibly lie in that (1) Pulkovo atmospheric refraction table is a global one and it doesn't have local characteristics which could be more prominent at larger zenith distance. The discrepancy between the measured and the tabular values may reflect the unconformity between Pulkovo table itself and the local actual atmospheric refraction; (2) Atmospheric refraction is related closely to environmental temperature and pressure. Subject to experiment conditions, the meteorological data including temperature and pressure cannot be recorded accurately in real-time. The recording errors of the meteorological data may bring some systematic difference between measured values and Pulkovo tabular values.

## 5 CONCLUDING REMARKS

It has always been considered difficult to measure the atmospheric refraction at lower angles of elevation. A new idea for determining atmospheric refraction by utilizing differential measurement with a telescope with double fields of view is proposed based on original work. A schematic prototype of atmospheric refraction with double fields of view was developed by the modification on a Maksutov tube. Experimental observations were carried out and the atmospheric refractions at lower angles of elevation can be obtained by the schematic prototype. Experiment results prove the feasibility of differential measurement of atmospheric refraction with double fields of view.

Table 3  Measured values of atmospheric refraction at different
observed zenith distances and comparison with Pulkovo table

| No. | Observed zenith distance | Number of observations | Measured values | Pul. Table values | Measured-Pul. table |
|-----|--------------------------|------------------------|-----------------|-------------------|---------------------|
| 1 | 49.4019566″ | 15 | 57.63″ ± 0.14″ | 57.68″ | -0.05″ |
| 2 | 63.9007239″ | 15 | 100.28″ ± 0.10″ | 100.60″ | -0.32″ |
| 3 | 69.0029459″ | 15 | 127.06″ ± 0.13″ | 128.04″ | -0.98″ |
| 4 | 74.0948008″ | 12 | 169.90″ ± 0.16″ | 171.46″ | -1.56″ |
| 5 | 78.8301660″ | 9 | 240.23″ ± 0.27″ | 244.11″ | -3.88″ |

However, we also found some shortcomings with the schematic prototype in experimental observations. Improvements are needed in the following aspects:

(1) Limited by the aperture and focal length of the Maksutov tube, the ability of gathering light is inadequate. It is hard for the schematic prototype to observe sufficient stars at lower angle of elevation where there is serious atmospheric extinction. The choice of telescope tube should be considered comprehensively from detective ability and observation accuracy.

(2) Meteorological recording instrument needs to be equipped with accurate record of temperature, humidity and barometric pressure to improve the accuracy of measuring atmospheric refraction. In addition, the variation of atmospheric refraction as a function of environmental factors can be studied on the basis of long-tern measurement.

(3) The automatic level of the schematic prototype is lower and manual intervention exists in observation. The automatic level of image processing and data reduction also needs improving further.

Acknowledgements  This work is supported by the National Natural Science Foundation of China (U1331112). We would like to acknowledge the assistance of Guo-Tao Gao at Anji station during the observations.